\documentclass[onecolumn,showpacs,preprintnumbers,amsmath,amssymb]{revtex4}
\usepackage{amsfonts,mathrsfs,graphicx}
\usepackage{epsfig}
\usepackage{bm}
\usepackage{color}
\def\be{\begin{equation}}
\def\ee{\end{equation}}
\def\bea{\begin{eqnarray}}
\def\eea{\end{eqnarray}}
\def\nnb{\nonumber}

\begin{document}

\title{THREE NEUTRINO OSCILLATIONS IN MATTER}

\author{
 Ara  Ioannisian$^{1,2}$  and Stefan Pokorski$^3$
 }
 \address{
 $^1$
Yerevan Physics Institute, Alikhanian Br.\ 2, 375036 Yerevan,
Armenia\\
$^2$Institute for Theoretical Physics and Modeling, 375036
Yerevan, Armenia\\
$^3$
Institute of Theoretical Physics, Faculty of Physics, University of Warsaw, ul. Pasteura 5, PL-02-093 Warsaw, Poland
}

\begin{abstract}
Following similar approaches in the past, the Schrodinger equation for  three neutrino propagation in matter of constant density is solved analytically by two successive  diagonalizations of  2x2 matrices. The final result for the oscillation probabilities is obtained  directly in the conventional parametric form as in the vacuum  but  with explicit simple modification of  two mixing angles ($\theta_{12}$ and $\theta_{13}$)  and mass eigenvalues. In this form, the analytical  results provide excellent approximation to  numerical calculations and allow for simple qualitative understanding of the matter effects.

\end{abstract}

%\preprint{CERN-TH-2018-256}
 
%\pacs{14.60.Pq, 26.65.+t, 91.35.-x,95.85.Ry, 96.60.Jw, }

\maketitle

The MSW effect 
~\cite{MSW} 
for  the neutrino propagation in matter attracts a lot of  experimental and theoretical attention. 
%{\bf WHY?} 
Most recently, the discussion is focused on the DUNE experiment 
~\cite{DUNE}.

%{\bf  Here a brief description of the Dune experiment, its parameters.}

On the theoretical  side, a large number of numerical simulations of the MSW effect in matter with  a constant or varying density has been performed.  Although, in principle, sufficient for comparing the theory predictions with experimental data, they do not provide a transparent physical  interpretation
of the experimental results. Therefore, several authors have also published analytical or semi-analytical solutions  to the  Schroedinger equation for three neutrino propagation in matter of constant density, in various perturbative expansions ~\cite{ many papers, Blennow:2013rca,  Denton:2016wmg}.
The complexity of  the calculation, the transparency of the final result and the range of its applicability depend on the chosen expansion parameter. 
%Such solutions do not exist at all for non-constant matter density.

In  this short note  we solve the Schroedinger equation in matter with constant density, using the approximate  see-saw structure of the full Hamiltonian  in the  electroweak basis.  This way one can diagonalize the 3x3 matrix by two successive diagonalizations of 2x2 matrices (similar approaches have been used in the past, in particular in ref. \cite{Blennow:2013rca} and \cite{Denton:2016wmg}). We specifically have in mind the parameters of the DUNE
experiment but our method is applicable for their  much wider  range.  The final result for the oscillation probabilities is obtained directly in  the conventional parametric form as in the vacuum but with modified  two mixing angles and mass eigenvalues\footnote {The results of this paper have been  presented as private communication by one of us (A.I) to the members of the T2HKK collaboration  in December 2017.}, similarly to the well  known results for the two-neutrino propagation in matter.
The three neutrino oscillation probabilities in matter have been presented in the same form  as here in the  recent ref.~\cite{PARKE ET AL}, where  the earlier results obtained in ref.
\cite{Denton:2016wmg}  are rewritten in this form. The form of our final results  can also  be obtained after some simplifications from ref.\cite{Blennow:2013rca}.
%similar results have been recently obtained in ref.
Our approach can be easily  generalized to  non-constant matter density by dividing the path of the neutrino trajectory in the matter to layers and assuming constant density in each layer.

%\section{Constant matter density}
 
The starting point is the Schroedinger equation
\be
i {d \over dx}\nu  = {\cal H}\nu   
%\nnb 
\ee

where ${\cal H}$ is the Hamiltonian in matter.  In the electroweak basis it reads
\be
{\cal H}=U 
\left( \begin {array}{ccc}
 0&0&0\\ 
0&{\Delta m^2_\odot \over 2E}&0\\ 
0&0&{\Delta m^2_a \over 2E}
\end {array} \right)
U^\dagger +
\left( \begin {array}{ccc}
 V(x)&0&0\\ 
0&0&0\\ 
0&0&0
\end {array} \right)
%\nnb
\label{electroweak}
\ee
The matrix $U$ is the neutrino mixing matrix in the vacuum. The mass squared differences are defined as $\Delta m^2_\odot \equiv m^2_2-m^2_1$ ($\approx  7.5 \ 10^{-5} eV^2$) and 
$\Delta m^2_a \equiv m^2_3-m^2_1$ ($\approx  \pm 2.5 \ 10^{-3} eV^2$, positive sign is for normal mass ordering and negative sign for inverted one).
Here $V(x)$ is the neutrino weak  interaction potential energy $V=\sqrt{2} G_F N_e$  ($N_e$ is electron number density) and we take it in this section to be x-independent.   The neutrino oscillation probabilities are determined by the $S$-matrix elements
\bea
S_{\alpha \beta}
=T \ e^{-i \int_{x_0}^{x_f} {\cal H}(x) dx}
\eea

For a  constant V  and in order to obtain our results  in  the same form as for the oscillation probabilities in the vacuum, it is convenient   to rewrite the $S$-matrix elements as follows:
\bea
S_{\alpha \beta}
=e^{-iU_m{\cal H}_mU^\dagger_m(x_f-x_0)}=U_me^{-i{\cal H}_mL}U^\dagger_m
\eea
The matrix $ {\cal H}_m$ is the Hamiltonian in matter in the mass eigenstate basis:
\bea
{
%\tiny
\left( 
\begin{tabular}{ccc}
${\cal H}_1$&0&0
\\
0&${\cal H}_2$&0
\\
0&0&${\cal H}_3$
\end{tabular} 
\right)
}
\eea
and the $U_m$ is the neutrino mixing matrix in matter.  Defining $\phi_{21}=({\cal H}_2-{\cal H}_1)L$
and $\phi_{31}=({\cal H}_3-{\cal H}_1)L$, we can write
\bea
&&
S_{\alpha \beta}
=\ \left[ U_m
\left( \begin {array}{ccc}
 1&0&0\\ 
0&e^{-i\phi_{21}}&0\\ 
0&0&e^{-i\phi_{31}}
\end {array} \right)
 U_m^\dagger \right] 
 \label{smatrix1}
\eea
 Here we neglect irrelevant overall phase, $e^{-i{\cal H} _1L}$. The neutrino transition probabilities do not depend on  the overall phase of the $S$ matrix. 

The remaining task is to find the eigenvalues  of ${\cal H}$ and the  mixing matrix $U_m$:
\be
{\cal H}= U_m{\cal H}_mU^\dagger_m
\ee
It is convenient to do it in two steps, first calculating the hamiltonian in a certain  auxiliary basis. This way, to an excellent approximation, we can diagonalize the 3x3 matrix by two successive diagonalizations of the 2x2 matrices.

The auxiliary basis \cite{Krastev:1988yu,Peres:2003wd} is defined  by  the following equation
\bea
{\cal H^\prime}=U^{aux\dagger}{\cal H} U^{aux} ~~and ~~S=U^{aux}e^{(-i{\cal H}^\prime L)} U^{aux\dagger}
\label{Hprime}
\eea
where
\bea
U^{aux}={\cal O}_{23}U^\delta{\cal O}_{13}
\eea
and the rotations ${\cal O}_{ij}$ are defined by the decomposition of
the mixing matrix $U$ in the vacuum (see eq.~\ref{electroweak})  as follows:
\bea 
\hspace{-1cm}
&&U={\cal O}_{23} U^\delta {\cal O}_{13} U^{\delta *} {\cal O}_{12} 
\nnb
\\ 
\hspace{-1cm}
&& \hspace{-0.5cm}
=
{
\small
\left( \! \! \! \!
\begin{tabular}{ccc}
$c_{13} c_{12}$ & $c_{13} s_{12}$ & $s_{13} e^{-i\delta}$ \\
$-s_{12} c_{23} - c_{12} s_{23} s_{13} e^{i\delta}$ &
   $c_{12} c_{23}-s_{12}s_{23}s_{13}e^{i\delta}$ & $c_{13}s_{23}$\\
$s_{12}s_{23}-c_{12}c_{23} s_{13} e^{i\delta}$ &
   $-c_{12} s_{23} - s_{12}c_{23} s_{13} e^{i\delta}$ &
         $c_{13}c_{23}$
\end{tabular} \! \! \!
\right)
}
\eea
where
\be
U^\delta =
\small
\left( \begin {array}{ccc}
 1&0&0\\ 
0&1&0\\ 
0&0&e^{i\delta}
\end {array} \right)
\ee
($c_{12}\equiv \cos \theta_{12}$, $s_{12}\equiv \sin \theta_{12}$ etc).

The matrices ${\cal O}_{ij}$ are orthogonal matrices. It is more convenient to rewrite the matrix $U$ in another form
\bea
&&
U \to \tilde{U}=U \cdot U^\delta
={\cal O}_{23} U^\delta {\cal O}_{13} {\cal O}_{12}
\nnb
\\
&&
\hspace{-1cm}
= \! \! \! \! 
\small
\left( \! \! \! \!
\begin{tabular}{ccc}
$c_{13} c_{12}$ & $c_{13} s_{12}$ & $s_{13}$ \\
$-s_{12} c_{23} - c_{12} s_{23} s_{13} e^{i\delta}$ &
   $c_{12} c_{23}-s_{12}s_{23}s_{13}e^{i\delta}$ & $c_{13}s_{23} e^{i\delta}$\\
$s_{12}s_{23}-c_{12}c_{23} s_{13} e^{i\delta}$ &
   $-c_{12} s_{23} - s_{12}c_{23} s_{13} e^{i\delta}$ &
         $c_{13}c_{23} e^{i\delta}$
\end{tabular} \! \! \!
\right)
\eea

Using eqs.~(\ref{electroweak},\ref{Hprime}) we obtain
\bea
{\cal H}^\prime &=&{\cal O}_{13}^T U^{\delta *} {\cal O}_{23}^T  \ {\cal H} \  {\cal O}_{23} U^\delta {\cal O}_{13}
\nnb
\\
&=& 
\small \! \! \!
\left(
\begin{tabular}{ccc}
$V c^2_{13} $ & $ s_{12} c_{12}{\Delta m^2_\odot \over 2E} $ &  $s_{13} c_{13}$ V\\
$ s_{12} c_{12}{\Delta m^2_\odot \over 2E} $  & $(c_{12}^2-s_{12}^2){\Delta m^2_\odot \over 2E} $ & 0 \\
$s_{13} c_{13}$ V& 0 &   ${\Delta m^2_{ee} \over 2E} + V s^2_{13}$
\end{tabular}
\right), 
\\
\Delta m^2_{ee}\!  \! &=&\! \!  c_{12}^2 \Delta m^2_{a}+ s_{12}^2 (\Delta m^2_{a}-\Delta m^2_\odot) 
%= \Delta m^2_{a} - s_{12}^2 \Delta m^2_\odot
\eea

The term $s_{12}^2\frac{\Delta m^2_\odot}{2E}$ has been subtracted from the diagonal elements; it gives an overall phase to the S-matrix  and according to the comments after eq. (\ref{smatrix1}) is irrelevant.

The definition of $\Delta m^2_{ee}$ coincides with one of the definitions of the effective mass squared differences measured at reactor experiments \cite{Patrignani:2016xqp,Nunokawa:2005nx}

This matrix has a see-saw structure, with the  $(13), (31)$ elements much smaller than  the $(33)$
element  and can be put in an almost diagonal form by two rotations
\bea
&& \hspace{-1cm}
{\cal O}_{12}^{m \ T}{\cal O}_{13}^{\prime \ T} {\cal H}^\prime \ {\cal O}_{13}^\prime  {\cal O}_{12}^m = 
{
\small
\left( 
\begin{tabular}{ccc}
${\cal H}_1$&0&0
\\
0&${\cal H}_2$&0
\\
0&0&${\cal H}_3$
\end{tabular} 
\right)
}
\\
&\equiv&
{
\small
\left( 
\begin{tabular}{ccc}
0&0&0
\\
0&${\Delta m^2_{21} \over 2E}$&0
\\
0&0&${\Delta m^2 _{31} \over 2E}$
\end{tabular} 
\right)
}+{\cal H}_1 {
%\tiny
\left( 
\begin{tabular}{ccc}
1&0&0
\\
0&1&0
\\
0&0&1
\end{tabular} 
\right)
}
\eea
After the first rotation we have
\begin{widetext}
\be
{\cal O}_{13}^{\prime T} {\cal H}^\prime   {\cal O}_{13}^{\prime}  =
{\small
\left(
\begin{tabular}{lll}
$\sin^2 \theta^\prime_{13}{\Delta m^2_{ee} \over 2E}+\cos^2 (\theta_{13}+\theta^\prime_{13}) V $ &  $\cos \theta^\prime_{13} \ s_{12} c_{12} {\Delta m^2_\odot \over 2E} $&0\\
$ \cos \theta^\prime_{13} \ s_{12} c_{12}{\Delta m^2_\odot \over 2E} $  & $(c_{12}^2-s_{12}^2){\Delta m^2_\odot \over 2E} $ & $ \sin \theta^\prime_{13} \ s_{12} c_{12}{\Delta m^2_\odot \over 2E} $ \\
0& $ \sin \theta^\prime_{13} \ s_{12} c_{12}{\Delta m^2_\odot \over 2E} $ & $\cos^2 \theta^\prime_{13}{\Delta m^2_{ee} \over 2E} + \sin^2 (\theta_{13}+\theta_{13}^\prime)V $
\end{tabular}
\right)}
\label{afr}
\ee
\end{widetext}
where

\bea
\sin 2 \theta_{13}^\prime &=& {\epsilon_a \sin 2 \theta_{13} \over \sqrt{(\cos 2\theta_{13}-\epsilon_a)^2+\sin^2 2 \theta_{13}}} , 
\eea
and
\bea
\epsilon_a &=& {2 E V \over \Delta m_{ee}^2}
\eea
We can safely neglect  the $(23), (32)$ elements  which are generated after the first rotation (see  Appendix A) and diagonalize the remaining 2x2 sub-matrix with the second rotation 
%\begin{widetext}
%\be
%{\cal O}_{13}^{\prime T} {\cal H}^\prime   {\cal O}_{13}^{\prime} \approx
%{\small
%\left(
%\begin{tabular}{lll}
%$\sin^2 \theta^\prime_{13}{\Delta m^2_{ee} \over 2E}+\cos^2 (\theta_{13}+\theta^\prime_{13}) V $ &  $\cos \theta^\prime_{13} \ s_{12} c_{12} {\Delta m^2_\odot \over 2E} $&0\\
%$ \cos \theta^\prime_{13} \ s_{12} c_{12}{\Delta m^2_\odot \over 2E} $  & $(c_{12}^2-s_{12}^2){\Delta m^2_\odot \over 2E} $ & 0 \\
%0& 0 & $\cos^2 \theta^\prime_{13}{\Delta m^2_{ee} \over 2E} + \sin^2 (\theta_{13}+\theta_{13}^\prime)V $
%\end{tabular}
%\right)}
%\ee
%\end{widetext}
%Now we can diagonalize the latter with the second rotation
\be  \
\sin 2 \theta_{12}^m = {\cos \theta^\prime_{13} \sin 2 \theta_{12} \over  \sqrt{(\cos 2 \theta_{12}-\epsilon_\odot)^2+\cos^2 \theta^{\prime}_{13} \sin^2 2 \theta_{12}} } , \ \ \ \sf where \ \ \ \  
\epsilon_\odot ={2EV \over \Delta m^2_\odot}(\cos^2 (\theta_{13}+ \theta_{13}^\prime) +{\sin^2 \theta^\prime_{13}\over \epsilon_a}) \ .
\ee
The eigenvalues of $\cal H$ are
\bea
\label{m21}
{\cal H}_2 - {\cal H}_1  \equiv  
{\Delta m^2_{21} \over 2 E}&=& {\Delta m_\odot^2 \over 2E} \sqrt{(\cos 2 \theta_{12}-\epsilon_\odot)^2+\cos^2 \theta^\prime_{13} \sin^2 2 \theta_{12}},
\\
\label{m31a}
{\cal H}_3- {\cal H}_1  \equiv  
{\Delta m^2_{31} \over 2 E}&=&\cos^2 \theta^\prime_{13}{\Delta m^2_{ee} \over 2E} + \sin^2 (\theta_{13}+\theta_{13}^\prime)V 
\nnb
\\
&& -{1 \over 2}[ (c_{12}^2-s_{12}^2){\Delta m^2_\odot \over 2E} +  \sin^2 \theta^\prime_{13}{\Delta m^2_{ee} \over 2E}+\cos^2 (\theta_{13}+\theta^\prime_{13}) V ]
\nnb
\\
&&+{1 \over 2}{\Delta m^2_{21} \over 2 E}
\\
%&=&{\Delta m^2_{ee} \over 4 E} \sqrt{(\cos 2\theta_{13}-\epsilon_a)^2+\sin^2 2 \theta_{13}} +{1 \over 2}[{\Delta m^2_{ee} \over 2 E} +V]
%\nnb
%\\
%&& -{1 \over 2}[ (c_{12}^2-s_{12}^2){\Delta m^2_\odot \over 2E} - {\Delta m^2_{ee} \over 4 E} \sqrt{(\cos 2\theta_{13}-\epsilon_a)^2+\sin^2 2 \theta_{13}} +{1 \over 2}[{\Delta m^2_{ee} \over 2 E} +V]
%\nnb
%\\
%&&+{1 \over 2}{\Delta m^2_{21} \over 2 E}
%\\
&=&{\Delta m^2_{ee} \over 2E} \sqrt{(\cos 2\theta_{13}-\epsilon_a)^2+\sin^2 2 \theta_{13}}
\nnb
\\
&& -{1 \over 4}{\Delta m^2_{ee} \over 2E} \sqrt{(\cos 2\theta_{13}-\epsilon_a)^2+\sin^2 2 \theta_{13}}+{1 \over 4}[{\Delta m^2_{ee} \over 2 E} +V]+{1 \over 4 E}(\Delta m^2_{21}-\Delta m^2_{\odot}\cos 2 \theta_{12})
\label{m31b}
%\nnb
\eea
Finally, for the mixing matrix in matter we obtain
\be
\tilde U_m = U^{aux}{\cal O}^\prime_{13}{\cal O}^m_{12}= {\cal O}_{23} U^\delta {\cal O}_{13}  {\cal O}_{13}^\prime {\cal O}_{12}^m  ={\cal O}_{23} U^\delta {\cal O}_{13}^m {\cal O}_{12}^m ,
%\nnb
\label{our}
\ee
For the matrix $U$ defined in eq.(10) we get
%\footnote{\textcolor{blue}{
%We want to underline that eq. (\ref{our}) and eq. (\ref{common}) are equivalent representations of the mixing matrix. $U^{\delta *}$ commutes with ${\cal O}_{12}^m$ and that phase %can be absobed by the neutrino wave functions. }
%}
\bea
U_m={\cal O}_{23}
\left( \begin {array}{ccc}
 1&0&0\\ 
0&1&0\\ 
0&0&e^{i\delta}
\end {array} \right)
\left( \begin {array}{rrr}
 \cos \theta^m_{13}& 0&\sin \theta^m_{13}\\
 0&1&0\\ 
 -\sin \theta^m_{13}& 0&\cos \theta^m_{13}
\end {array} \right)
\left( \begin {array}{ccc}
 1&0&0\\ 
0&1&0\\ 
0&0&e^{-i\delta}
\end {array} \right)
\left( \begin {array}{rrr}
 \cos \theta^m_{12}& \sin \theta^m_{12}&0\\ 
 -\sin \theta^m_{12}& \cos \theta^m_{12}&0\\ 
   0&0&1
\end {array} \right)
%\nnb
\label{common}
\eea

with $\theta_{13}^m =\theta_{13}+ \theta_{13}^\prime$ and 
\be
\sin 2 \theta_{13}^m = {\sin 2 \theta_{13} \over \sqrt{(\cos 2\theta_{13}-\epsilon_a)^2+\sin^2 2 \theta_{13}}} , 
\hspace{1.6cm}
\cos 2 \theta_{13}^m \ = \ {\cos 2\theta_{13}-\epsilon_a \over \sqrt{(\cos 2\theta_{13}-\epsilon_a)^2+\sin^2 2 \theta_{13}}} 
\label{t13}
\ee
and
\be
\sin 2 \theta_{12}^m = {\cos \theta^\prime_{13} \sin 2 \theta_{12} \over  \sqrt{(\cos 2 \theta_{12}-\epsilon_\odot)^2+\cos^2 \theta^{\prime}_{13} \sin^2 2 \theta_{12}} }, 
\ \ \
\cos 2 \theta_{12}^m  \ = \  { \cos 2 \theta_{12}-\epsilon_\odot \over  \sqrt{(\cos 2 \theta_{12}-\epsilon_\odot)^2+\cos^2 \theta^{\prime}_{13} \sin^2 2 \theta_{12}} }
\label{t12}
\ee

%\bea
%{\Delta m^2_{21} \over 2 E}&=& {\Delta m_\odot^2 \over 2E} \sqrt{(\cos 2 %\theta_{12}-\epsilon_\odot)^2+\cos^2 \theta^\prime_{13} \sin^2 2 \theta_{12}},
%\\
%{\Delta m^2_{31} \over 2 E}&=&{\Delta m^2_{ee} \over 2E}  +{1 \over 2}({\Delta m^2_{21}  \over %2E}-{\Delta m^2_\odot  \over 2E} \cos 2 \theta_{12}-\cos^2 \theta_{13}^m V) + \sin^2 %\theta^m_{13}V - {3 \over 2}\sin^2 \theta^\prime_{13}{\Delta m^2_{ee} \over 2E}
%\\
%&=&{\Delta m^2_{ee} \over 2E} \sqrt{(\cos 2\theta_{13}-\epsilon_a)^2+\sin^2 2 \theta_{13}}
%\nnb
%\\
%&& -{1 \over 4}{\Delta m^2_{ee} \over 2E} \sqrt{(\cos 2\theta_{13}-\epsilon_a)^2+\sin^2 2 %\theta_{13}}+{1 \over 4}[{\Delta m^2_{ee} \over 2 E} +V]+{1 \over 4 E}(\Delta m^2_{21}-\Delta %m^2_{\odot}\cos 2 \theta_{12})
%\\
%\epsilon_\odot &=& {2EV \over \Delta m^2_\odot}(\cos^2 \theta_{13}^m +{\sin^2 %\theta^\prime_{13}\over \epsilon_a})
%\eea

{\bf In summary}  the 
mixing matrix in matter, $U_m$, is given by  the following change  of the parameters  from the vacuum solution:

$\theta_{12} \to \theta_{12}^m$ \ \ \ (eq,\ref{t12})

$\theta_{13} \to \theta_{13}^m$ \ \ \ (eq.\ref{t13})

$\theta_{23}^m \equiv \theta_{23}$

$\delta^m \equiv \delta$.

The mass eigenvalues are  given by eqs.\ref{m21},\ref{m31a},\ref{m31b}.

%{\Delta m^2_{21} \over 2 E}&=& {\Delta m_\odot^2 \over 2E} \sqrt{(\cos 2 %\theta_{12}-\epsilon_\odot)^2+\cos^2 \theta^\prime_{13} \sin^2 2 \theta_{12}},
%\\
%{\Delta m^2_{31} \over 2 E}&=&{\Delta m^2_{ee} \over 2E}  +{1 \over 2}({\Delta m^2_{21}  \over %2E}-{\Delta m^2_\odot  \over 2E} \cos 2 \theta_{12}-\cos^2 \theta_{13}^m V) + \sin^2 %\theta^m_{13}V - {3 \over 2}\sin^2 \theta^\prime_{13}{\Delta m^2_{ee} \over 2E}
%\\
%&=&{\Delta m^2_{ee} \over 2E} \sqrt{(\cos 2\theta_{13}-\epsilon_a)^2+\sin^2 2 \theta_{13}}
%\nnb
%\\
%&& -{1 \over 4}{\Delta m^2_{ee} \over 2E} \sqrt{(\cos 2\theta_{13}-\epsilon_a)^2+\sin^2 2 %\theta_{13}}+{1 \over 4}[{\Delta m^2_{ee} \over 2 E} +V]+{1 \over 4 E}(\Delta m^2_{21}-\Delta %m^2_{\odot}\cos 2 \theta_{12})
%\\
%\nnb
%\\
%\theta_{13}^m&=&\theta_{13}+ \theta_{13}^\prime 
%\\
%\epsilon_\odot &=& {2EV \over \Delta m^2_\odot}(\cos^2 \theta_{13}^m +{\sin^2 %\theta^\prime_{13}\over \epsilon_a}) \ , 
%\ \ \ \ \ \ \
%\epsilon_a = {2EV \over \Delta m^2_{ee}}
%\\
%\Delta m^2_{ee}&=& c_{12}^2 \Delta m^2_{a}+ s_{12}^2 (\Delta m^2_{a}-\Delta m^2_\odot)= \Delta %m^2_{a}- s_{12}^2 \Delta m^2_\odot
%\eea

The oscillation probabilities $P_{\nu_\alpha \to \nu_\beta}$  ($\alpha, \beta = e, \mu, \tau$)  have the same forms as for the  vacuum oscillations with mass eigenstates as  above and with replacements $\theta_{12} \to \theta_{12} ^m $ and  $\theta_{13} \to \theta_{13} ^m $. For the $\nu_\mu\rightarrow \nu_e$ transition we have

\bea
P_{\nu_\mu \to \nu_e} &=& \sin^2 2 \theta^m_{13} s_{23}^2 \  \left[ c_{12}^{m2} \sin^2 {\phi_{31} \over 2}+ s_{12}^{m2} \sin^2 {\phi_{32} \over 2}\right]
\nnb
\\
&&
+{1 \over 2}c_{13}^m \sin 2 \theta^m_{13} \sin 2 \theta^m_{12} \sin 2 \theta_{23} \cos \delta \ \sin {\phi_{21} \over 2} \sin {\phi_{31} + \phi_{32}  \over 2} 
\nnb
\\
&&
- c_{13}^m \sin 2 \theta^m_{13} \sin 2 \theta^m_{12} \sin 2 \theta_{23} \sin \delta \ \sin {\phi_{21} \over 2} \sin {\phi_{31}  \over 2} \sin {\phi_{32}   \over 2} 
\nnb\\
&&
+\left[  c_{13}^{m2} \sin^2 2 \theta^m_{12} (c_{23}^2-s_{23}^2 s_{13}^{m2}) +{1 \over 4}c^m_{13} \sin 2\theta^m_{13}\sin 4\theta^m_{12}\sin 2\theta_{23} \cos \delta
 \right] \sin^2 {\phi_{21} \over 2}
 \label{Pmue}
\eea
where
\be
\phi_{ij} ={\Delta m^2_{ij} \over 2E}L \ \ \ \ i,j =1,2, 3 \  \ \ \ (L=1285 \ \text{km for  DUNE}) \ .
\ee

This  approximate solution is valid for all energies.  Numerically our result is identical to the approximation of  two angles rotation in \cite{Blennow:2013rca} and  the  0th order result of \cite{Denton:2016wmg}.
For  anti-neutrino oscillations $P_{{\bar \nu}_\alpha \to {\bar \nu}_\beta}$,  V$\to$ -V and $\delta \to -\delta$. 
For normal mass hierarchy $\Delta m_a^2$ is positive and for inverted mass hierarchy it is negative.

\begin{figure}[h!]
\includegraphics[width=1\textwidth, height=0.4\textwidth]{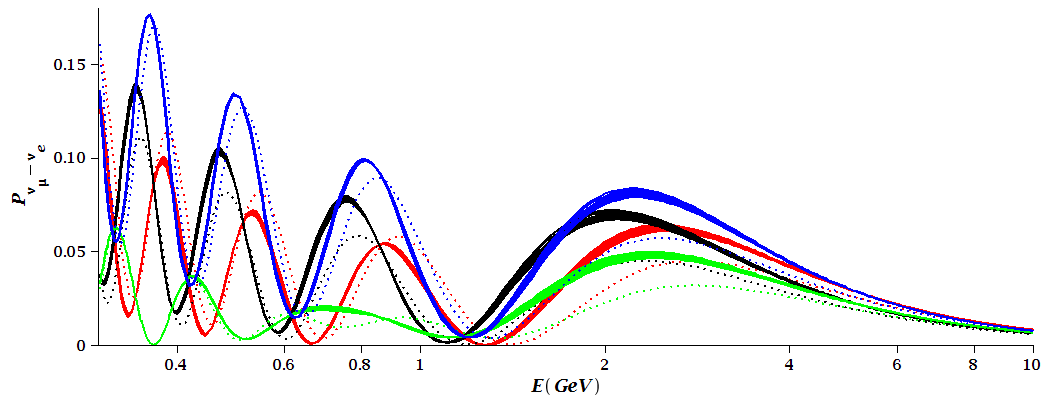}
\caption{$\nu_\mu \to \nu_e$ oscillation probability at DUNE for normal mass hierarchy, $\delta_{cp} = 0$ (red), $\delta_{cp} = {\pi \over 2}$ (green),  $\delta_{cp} = {\pi }$ (black), $\delta_{cp} = -{\pi \over 2}$ (blue). Thickness of the plots are from varying constant/uniform matter density 2.5 - 3 g/cm$^3$. Dotted plots are for  vacuum oscillations}
\label{NNH}. 
\end{figure}

\begin{figure}[h!]
\includegraphics[width=1\textwidth, height=0.4\textwidth]{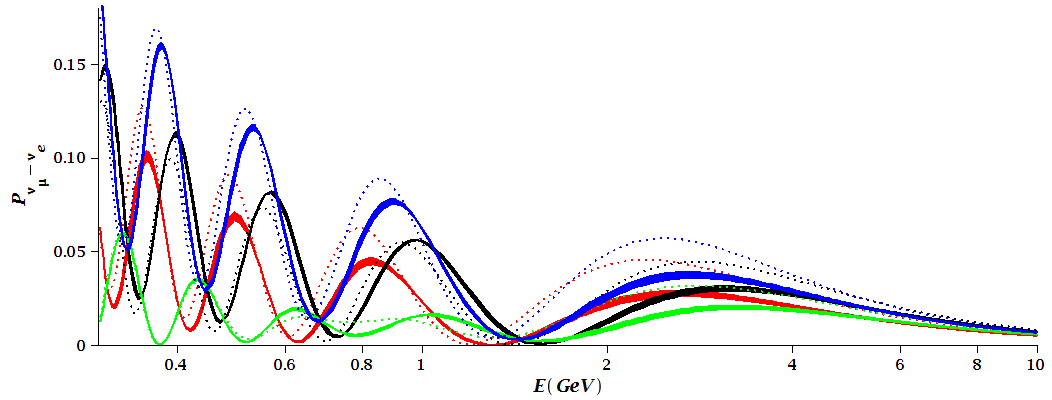}
\caption{$\nu_\mu \to \nu_e$ oscillation probability at DUNE for inverted mass hierarchy, $\delta_{cp} = 0$ (red), $\delta_{cp} = {\pi \over 2}$ (green), $\delta_{cp} = {\pi }$ (black), $\delta_{cp} = -{\pi \over 2}$ (blue). Thickness of the plots are from varying constant/uniform matter density 2.5 - 3 g/cm$^3$. Dotted plots are for  vacuum oscillations}
\label{NIH}. 
\end{figure}

Our solutions are illustrated in Fig.~\ref{NNH} and~\ref{NIH} for $\nu_\mu \to \nu_e$ oscillation at DUNE distance for several  values of $\delta_{CP}$ and compared with the oscillation probabilities in the vacuum, shown by the dotted curves. They are the reference point of our  discussion.

The matter effects and their dependence on $\delta_{CP}$ observed in those plots have easy explanation in terms of our analytic formulas. For the sake of definiteness, we focus on the region of the first maximum ($E=(1-6)$GeV), accessible in the DUNE experiment. 

First of  all we notice that
the modification in matter of the solar sector parameters, angle $\theta_{12}$  and $\Delta m^2_{21}$, have  very small effect when the phase $\phi_{21} \ll 1$ (as it is for DUNE distance and energies). In the first term of the rhs eq (\ref{Pmue}) the dependance on the $\phi_{21}$ is sub-leading. In the 2nd and 3rd terms we have combinations $\sin 2 \theta_{12}^m \sin {\phi_{21} \over 2 } $  and that for small phases can be rewritten as 
$$
 \sin 2 \theta_{12}^m \ \sin { \Delta m_{21}^2 \over 4E } L  \simeq  \sin 2 \theta_{12}^m \ { \Delta m_{21}^2 \over 4E } L= \cos \theta_{13}^\prime \sin 2 \theta_{12} \ { \Delta m_{\odot}^2 \over 4E } L \ .
 $$
which is almost independent on matter ($\cos \theta^\prime_{13} \simeq 1$ at $E=1-6$GeV). Therefore the dependence on matter due to change of  $\theta_{12}-\Delta m^2_{21}$ is very suppressed at DUNE.

The most important effect is the dependence of the oscillation  probability on  the angle $\theta_{13}$
which  has larger (smaller) values in matter than in the vacuum for normal (inverted)  neutrino mass hierarchies (and opposite for antineutrinos). Thus the oscillation probabilities have larger(lower) oscillation amplitudes for normal (inverted)  neutrino mass hierarchies (and opposite for antineutrinos). In oder words the matter of the Earth is amplifying the effect of the mass ordering on neutrino oscillations. The dependence on the angle $\theta_{13}$ enters multiplicatively in the first three terms of eq (\ref{Pmue}), whereas the fourth term is small in the region of the first maximum.
Therefore the matter effects relative to the oscillations in the vacuum do not depend on the value of $\delta_{CP}$, as it is seen in Fig.~\ref{NNH} and~\ref{NIH}. Moving to the next resonances (lower energies) the difference between
oscillations in matter and in the vaccuum remain qualitatively similar, although   some small differences can be seen due to the fact that the change in the angle $\theta_{13}$ is smaller.

 %(for $E\gg E_{res1}\simeq$ 300 MeV)

%In eq (\ref{Pmue}) the first two terms are CP even, the 3rd term is CP odd and the 4th term is very %small at first maximum ($\sin 2 \theta_{12}^m \ll 1$ at energies much above the first resonance). 

%For maximum CP violation cases ($\delta_{CP}=\pm \pi/2$) according to the 3rd term of rhs of the %eq (\ref{Pmue})
 %the amplitude of the neutrino oscillation probability with  $\delta_{CP}=- \pi/2$ is higher than with  %$\delta_{CP}= \pi/2$ (independent of neutrino mass ordering) and opposite for antineutrinos. 

%For CP conserved case  ($\delta_{CP}= 0, \pi$)  the second term in rhs  of eq (\ref{Pmue})  is positive %for normal ordering and it is negative for inverted ordering for both neutrinos and antineutrinos.  %That term is zero at the first maximum ($\phi_{31}\simeq \phi_{32}=\pm \pi /2$). Also due to the %twice times larger phase of that term compared to the first term (also third term) the maximums of %the lines with  $\delta_{CP}= 0, \pi$ are shifted in phase. 

\begin{figure}[h!]
\includegraphics[width=1\textwidth, height=0.4\textwidth]{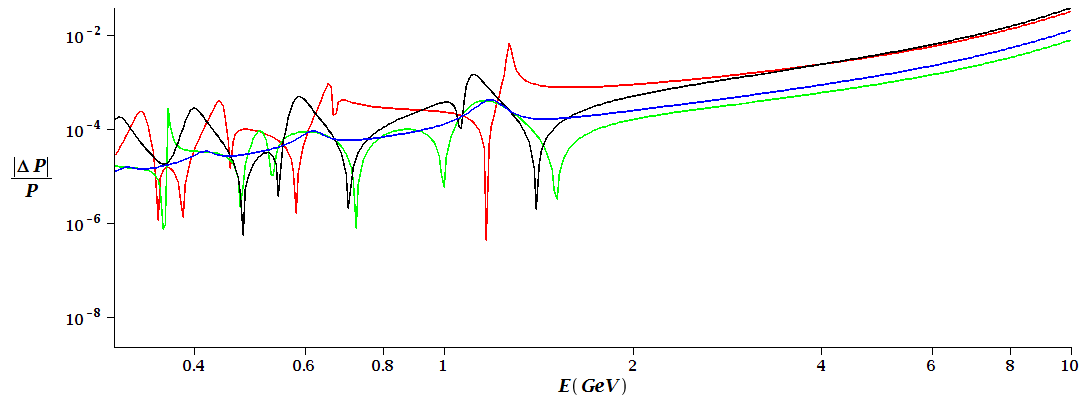}
\caption{${|\Delta P| \over P}  \equiv {|P^{num}_{\nu_\mu \to \nu_e}-P^{anl}_{\nu_\mu \to \nu_e}| \over P^{num}_{\nu_\mu \to \nu_e}}$. The relative error of our analytic result to the exact (numeric) $\nu_\mu \to \nu_e$ oscillation probability for normal mass hierarchy, $\delta_{cp} = 0$ (red), $\delta_{cp} = {\pi \over 2}$ (green), $\delta_{cp} = {\pi }$ (black), $\delta_{cp} = -{\pi \over 2}$ (blue). Matter density 2.6 g/cm$^3$. }
\label{RENN}. 
\end{figure}

Finally, in Fig.~\ref{RENN} we show the accuracy of the analytical solutions  comparing them with numerical/exact results.

%\end{document}

{\bf Appendix}

In ordinary perturbation expansion in the basis of our solutions we estimate the size of effect of the neglected elements $(23), (32)$ in eq. (\ref{afr}).

We divide the hamiltonian into two parts
\be
S= e^{-i{\cal H}L}=e^{-i{\cal H}_0 L-i\Delta {\cal H}L}
\ee
Here ${\cal H}_0 \equiv U_m diag(0,{\Delta m_{21}^2 \over 2E}L ,{\Delta m_{31}^2 \over 2E}L) \ U_m^\dagger$ with our solutions for $U_m$ (eq \ref{common}) and mass square differences (eqs. \ref{m21} -\ref{m31b}). $\Delta {\cal H}(23)=\Delta {\cal H}(32)=\sin \theta^\prime_{13} \sin 2\theta_{12} {\Delta m^2_\odot \over 4E}$ and all other elements in $\Delta {\cal H}$ are zeros. We treat  $\Delta {\cal H}$ as a perturbation.

By making use well known identity $e^{a+b}=e^a T e^{\int_0^1dt  \ e^{-a \ t} \ b \ e^{a \ t}}$ we get 

\bea
S&=&S_0+S_1+\dots
\\
S_0&=& 
U_m  \left( \begin {array}{ccc}
 1&0&0\\ 
0&e^{-i\phi_{21}}&0\\ 
0&0&e^{-i\phi_{31}}
\end {array} \right) 
U_m^\dagger \ \ , \ \ 
\\
S_1&=& 
U_m  \left( \begin {array}{ccc}
0&0&\cal A\\ 
0&0&\cal B\\ 
\cal A& \cal B&0
\end {array} \right) 
U_m^\dagger
\eea
where
\bea
\cal A &=&  \sin \theta^m_{12} \ \sin \theta^\prime_{13} {\sin 2 \theta_{12} \over 2} {\Delta m^2_\odot \over   \Delta m_{31}^2} (1 - e^{-i\phi_{31}})
\nnb
\\
\cal B &=&  \cos \theta^m_{12} \ \sin \theta^\prime_{13} {\sin 2 \theta_{12} \over 2} {\Delta m^2_\odot \over   \Delta m_{31}^2- \Delta m_{21}^2} (e^{-i\phi_{31}}-e^{-i\phi_{21}})
\nnb
%S_1&=& 
% \sin \theta^\prime_{13} {\sin 2 \theta_{12} \over 2} \Delta m^2_\odot \ 
%U_m
%\left(
%\begin{tabular}{ccc}
%0 & 0& $   -\sin \theta^m_{12}{e^{-i\phi_{31}}-1 \over  \Delta m_{31}^2}$ \\
%  0&  0 &$  \cos \theta^m_{12}{e^{-i\phi_{31}}-e^{-i\phi_{21}} \over \phi_{31}-\phi_{21}}$ \\
%$    -\sin \theta^m_{12}{e^{-i\phi_{31}}-1\over \Delta m_{31}^2}$  & $  \cos \theta^m_{12} {e^{-i\phi_{31}}-e^{-i\phi_{21}}\over \phi_{31}-\phi_{21}}$ &    $ 0$
%\end{tabular}
%\right) 
%U_m^\dagger
\eea

$S_0$ is our solution for the neutrino transition matrix elements and the $S_1$ is its first order corrections. 
$|\cal A|$, $|\cal B|$ are at least smaller than 0.5\% for  all energies, therefore our  0$^{th}$ order solution, $S_0$, is working excellently.

\vspace{0.3cm}

{\bf Acknowledgements.} One of us (A.I.) is grateful to the CERN Theory group for its hospitality and to the DUNE members for useful discussions at Fermilab in summer 2017. A.I. is especially grateful to Pilar Coloma and Maury Goodman for underlining the importance of presenting the oscillation parameters in a simple form.
S.P. is partially supported by the National Science
Centre, Poland, under research grants
% JR grant
DEC-2015/19/B/ST2/02848,
% Harmonia
DEC-2015/18/M/ST2/00054
% MO grant
and DEC-2014/15/B/ST2/02157.  

\end{document}